\documentclass{iau_FM}
\usepackage{graphicx}

\usepackage{cite}
\usepackage[round, sort]{natbib} 
\usepackage{color}
\usepackage[
colorlinks=true,	
linkcolor=blue,  
urlcolor=blue,   
citecolor=blue,  
filecolor=blue, anchorcolor=blue, runcolor=blue, 
hyperfootnotes=true
]{hyperref}											
\usepackage[hang, flushmargin]{footmisc}

\usepackage{ccaption}
\usepackage[footnotesize,labelfont=bf]{caption}  

\def \aap{\textit{A\&A}}

\def \apj{\textit{ApJ}}

\def\araa{\textit{ARAA}}
\def \mnras{\textit{MNRAS}}

\def \sci{\textit{Science}}
\def \nat{\textit{Nature}}

\def\msun{{\,M_\odot}}

\newcommand{\lsim}{\,\rlap{\raise 0.35ex\hbox{$<$}}{\lower 0.7ex\hbox{$\sim$}}\,}
\newcommand{\gsim}{\,\rlap{\raise 0.35ex\hbox{$>$}}{\lower 0.7ex\hbox{$\sim$}}\,}

\def\farcs{\hbox{$.\!\!^{\prime\prime}$}}

\title[The Radio Remnant of  Supernova 1987A] 
{The Radio Remnant of  Supernova 1987A $-$ \\
 A Broader View}

\author[Zanardo G., et al.]
{
G. Zanardo$^1$,
L. Staveley-Smith$^{1,2}$
C. -Y. Ng$^3$, 
R. Indebetouw$^{4,5}$, \\
M. Matsuura$^{6,7}$,
B. M. Gaensler$^{8,2}$,
A. K. Tzioumis$^9$
}

\affiliation{
$^1$International Centre for Radio Astronomy Research (ICRAR), \\ M468, The University of Western Australia, Crawley, WA 6009, Australia. \\ email: {\tt giovanna.zanardo@gmail.com} \\[\affilskip]
$^2$Australian Research Council Centre of Excellence for All-sky Astrophysics (CAASTRO) \\[\affilskip]
$^3$Department of Physics, The University of Hong Kong, Pokfulam Road, Hong Kong \\[\affilskip]
$^4$Department of Astronomy, University of Virginia, P.O. Box 400325, \\Charlottesville, VA 22904-4325, USA \\[\affilskip]
$^5$National Radio Astronomy Observatory (NRAO), 520 Edgemont Road, Charlottesville, VA 22903, USA \\[\affilskip]
$^6$School of Physics and Astronomy, Cardiff University, QueenÕs Buildings, The Parade, \\Cardiff, CF24 3AA, UK  \\[\affilskip]
$^7$Department of Physics and Astronomy, University College London, Gower Street, \\London, WC1E 6BT, UK \\[\affilskip]
$^8$Dunlap Institute for Astronomy \& Astrophysics, University of Toronto, \\Toronto, ON M5S 3H4, 
Canada  \\[\affilskip]
$^9$CSIRO Astronomy and Space Science, Australia Telescope National Facility, \\PO Box 76, Epping, NSW 1710, Australia \\[\affilskip]
}

\pubyear{2017}
\volume{331}  
\pagerange{119--126}
\jname{SN 1987A 30 years later}
\editors{M. Renaud, A. Marcowith, G. Dubner, A. Ray \& A. Bykov}
\begin{document}

\maketitle
\vspace{-2mm}
\begin{abstract}
Supernova remnants (SNRs) are powerful particle accelerators. As a supernova (SN) blast wave propagates through the circumstellar medium (CSM), electrons and protons scatter across the shock and gain energy by entrapment in the magnetic field. The accelerated particles generate further magnetic field fluctuations and local amplification, leading to cosmic ray production.
The wealth of data from Supernova 1987A is providing a template of the SN-CSM interaction, and an important guide to the radio detection and identification of core-collapse SNe based on their spectral properties. Thirty years after the explosion, radio observations of SNR 1987A span from 70 MHz to 700 GHz. 
We review extensive observing campaigns with the Australia Telescope Compact Array (ATCA)
and the Atacama Large Millimeter/submillimeter Array (ALMA), and follow-ups with other radio telescopes. 
Observations across the radio spectrum indicate rapid changes in the remnant morphology,
while current ATCA and ALMA observations show that the SNR has entered a new evolutionary phase. 

\end{abstract}

\firstsection 
\vspace{-2.0mm}              
\section{Background}

Radio supernovae are most consistently identified with core-collapse supernovae (CCSNe) and their transition into supernova remnants (SNRs).
The radio emission results from the collision of the supernova (SN) shock and the progenitor's circumstellar medium (CSM). 
Many radio SNe can be associated with H$\textsc{II}$  regions or regions of active star formation. Given the high ambient pressure of the surrounding H$\textsc{II}$ region, radio SNe and their remnants are primarily identified by dense CSM environments.
The SN shock wave modifies the CSM as it passes through it, both by compression and shock heating. The forward SN shock, i.e. the outer blast wave, and the reverse shock, which causes deceleration and heating of the inner ejecta, accelerate the shocked material, with some fractions of electrons and protons reaching cosmic ray energies. The acceleration mechanism usually associated with the shock propagation is diffusive shock acceleration (DSA), also known as the first-order Fermi mechanism \citep{dru83}. According to DSA theory, energetic particles at shock fronts undergo spatial diffusion in a uniform magnetic field. If the magnetic field is not uniform, as in the case of young SNRs, 
the charged particles can get partially trapped in the magnetic field structure, 
while still gaining energy by scattering between the upstream and downstream regions of the forward shock \citep{kir96}.
This leads to 
a sub-diffusive particle transfer (sub-DSA), 
since the particle density at the shock front is lower than it is far downstream. 
In either case, as energetic electrons gyrate in the local magnetic field, they emit synchrotron emission.
The 
synchrotron emission observed in SNRs 
can be well described by a power law $S_{\nu}\propto\nu^{\alpha}$, with $S_{\nu}$ the flux density measured at the observing frequency $\nu$. Simplified synchrotron physics tells us that emission at the frequency $\nu$ is primarily by electrons with energy 
$E \propto 15 \sqrt{\nu / B}$ GeV, with $\nu$ expressed in GHz and the magnetic field strength, $B$, expressed in $\mu$G \citep{rey11}.

SN 1987A  
in the Large Magellanic Cloud, 
identified as Type IIP \citep{lei03} with progenitor of mass $\sim 20 \msun$ \citep{sma09},
is the only nearby CCSN observed with a telescope since its early stages.
Given its proximity, SN 1987A has allowed unique studies of the radio evolution
of CCSNe and of the 
SN$-$CSM interaction.
The complex CSM distribution in the remnant is believed to have originated from a red supergiant (RSG) which has evolved into a blue supergiant (BSG) about 20,000 years before the explosion \citep{cro00}.
In its transition from supernova to SNR, SN 1987A has continued to brighten for the past three decades, thus making possible to monitor
the rapid changes of the radio emission with a number of telescopes over an increasingly broad
frequency range.

%
%
\begin{figure*}[!t]
\begin{center}
\vspace{0mm}
\advance\leftskip-0.0mm
\includegraphics[trim=0.11mm 0.09mm 0.0mm 0.15mm, clip=true,width=132.0mm, angle=0]{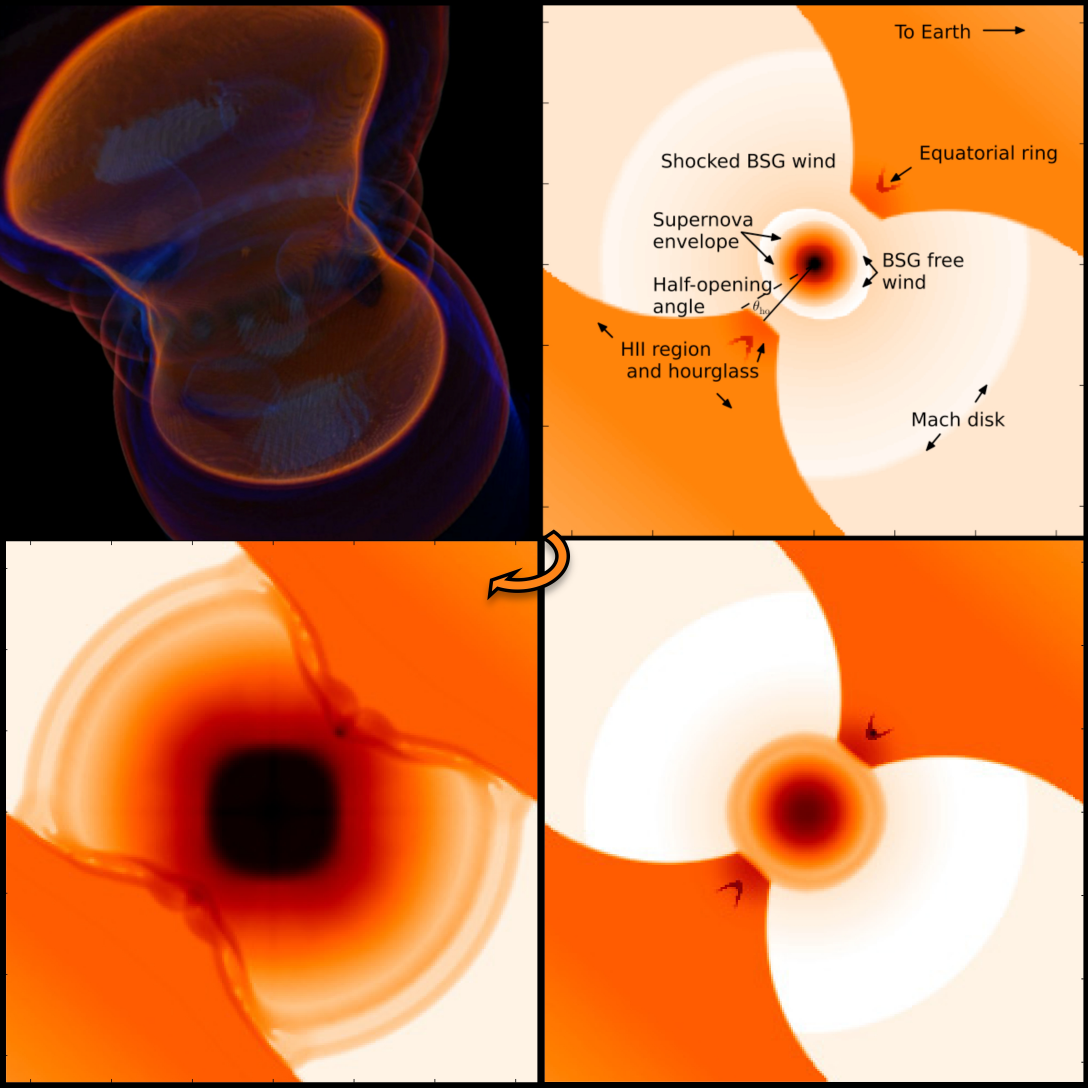}
\end{center}
\vspace{-1.0mm}
\caption{\footnotesize{
Stages of the shock expansion from HD simulations by \citet{pot14} (see their Figures 1, 6, and 7), as modelled around 1000 days (\textit{top right}), at day $\sim$2000 (\textit{bottom right}),
and at day $\sim$8000 (\textit{bottom left}) since the SN. By day $\sim$8000, the forward shock has  overtaken the equatorial ring (ER). The colour scheme varies from white to dark brown, indicating the increasing particle number density in log$_{10}$ scale ($\sim5-11\times10^{7}\, \,\textrm{m}^{-3}$, see \citealt{pot14}). A 3D view of both the forward and reverse shocks expanding above and below the ER, is shown on the top left image, which is a rendering of the volumetric models developed by \cite{pot14} (see their Figure 9) at day $\sim$8000. \\ }
\label{fig:hdmodels}}
\end{figure*}

\vspace{-3mm}
\section{Evolution of the SN-CSM interaction}
\label{SN_CSM}

The interaction of the SN shock wave with the CSM in the equatorial plane (equatorial ring, ER) is the most notable feature of the radio remnant of SN 1987A. The ER defines the waist of the `hourglass' environment created by the progenitor star prior to explosion. 
 As the SN propagates 
outwards and impacts  the denser CSM in the ER, the shocks expand above and below the ER plane (see Figure~\ref{fig:hdmodels}) 
 and interact with high-latitude material \citep{pot14}, confined within the inner hourglass structure.

\vspace{-1mm}
\subsection{Light curves} 
\label{light}
\begin{figure*}[!htb]
\begin{minipage}[c]{88mm}
\begin{center}
\vspace{-0mm}
\advance\leftskip-17.0mm
\includegraphics[width=88.0mm, angle=0]{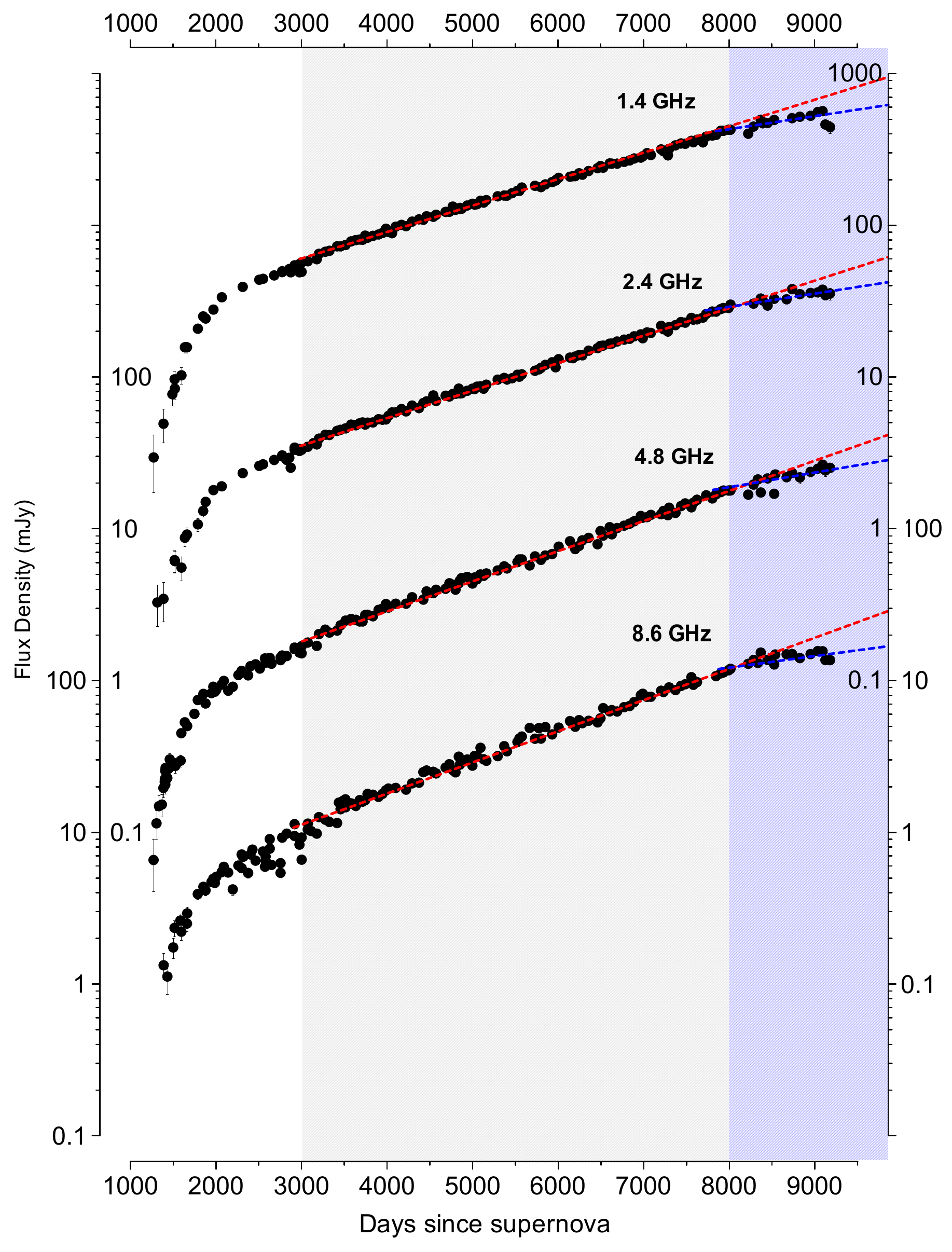}
\end{center}
\end{minipage}
\begin{minipage}[c]{88mm}
\begin{center}
\vspace{-3.8mm}
\advance\leftskip-49.5mm
\includegraphics[width=61.6mm, angle=0]{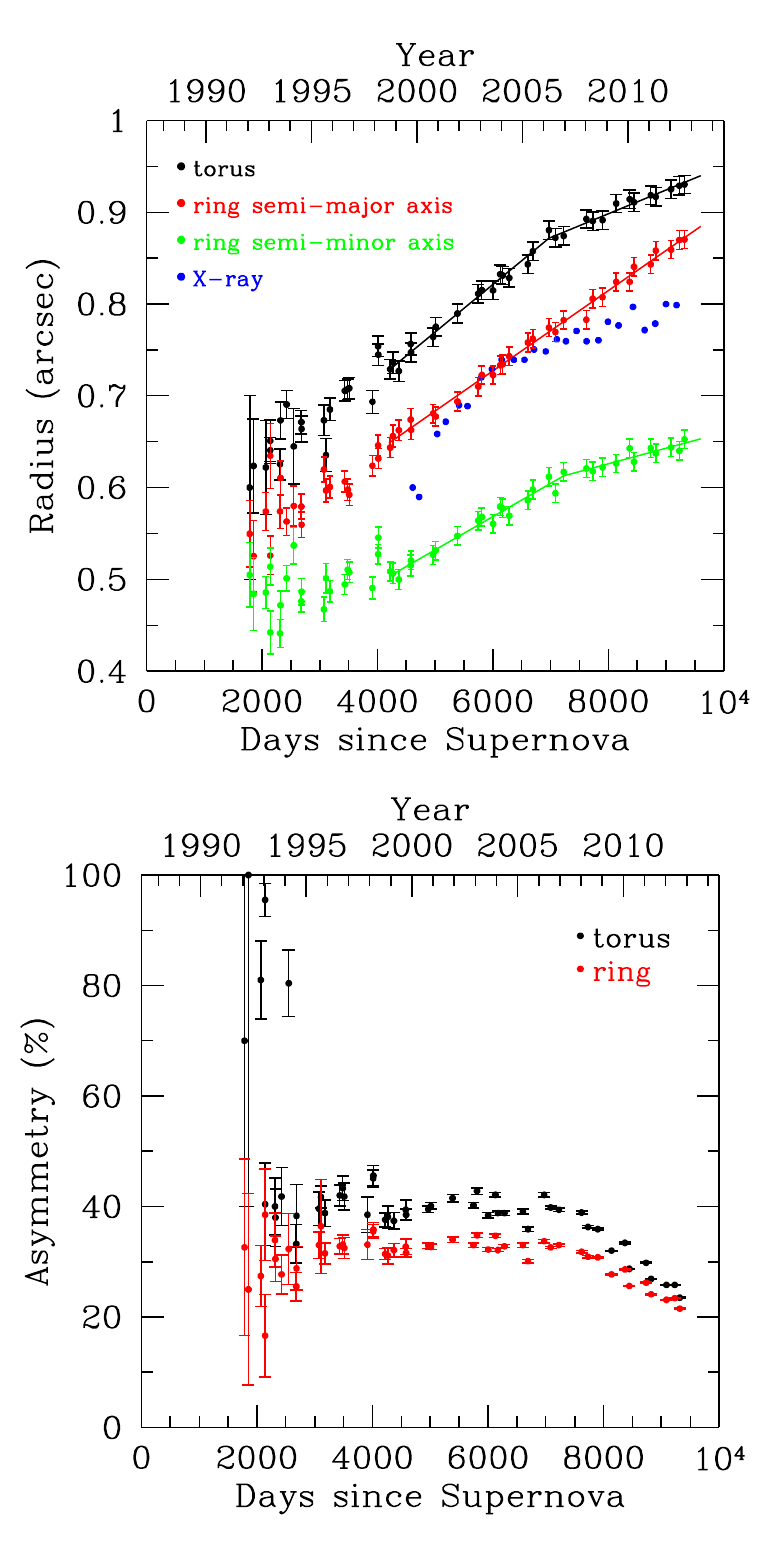}
\end{center}
\end{minipage}
\vspace{-2.0mm}
\caption{\small{{\it Left} $-$ Flux densities for SNR 1987A. Plots include data from observations at 1.4, 2.4, 4.8, and 8.6 GHz from August 1990 to June 2012 and are stacked with an offset of  0.1 mJy (\citealp{zan10,zan14b}). The red dashed lines are the exponential fits of data from November 2000 to February 2009, as derived by \cite{zan10}. The blue dashed lines are 1500-day fits of data from February 2009. The colour bands are included to highlight different rates of the flux density increase. {\it Top Right} $-$ Remnant radius resultant from images at 9 GHz, from August 1992 to March 2013, as derived by \cite{ng13} via fitting the SNR with a torus ({\it black}) and an elliptical ring ({\it red}: semi-major axis; {\it green}: semi-minor axis). The radius of the radio remnant is compared with that derived from X-ray images ({\it blue}, from \citealt{hel13}). The solid lines indicate the best-fit expansion. {\it Bottom Right} $-$ Evolution of the remnant asymmetry at  9 GHz as derived from the torus ({\it black}) and ring ({\it red}) models  \citep{ng13}. Credit: \citet{zan14b}, \citet{ng13}.}
\label{fig:monitoring}}
\end{figure*}

The radio emission from SN 1987A was first detected two days after the arrival of the neutrinos with the Molonglo Observatory Synthesis Telescope (MOST) at 843 MHz, 1.4, 2.3, and 8.4 GHz \citep{tur87}.
The peak flux density was reached at $\sim130$ mJy at 843 MHz four days after the explosion \citep{tur87},  followed by a power-law decay of the emission, which became undetectable from September 1987 \citep{bal95}. It is understood that this radio outburst was due to the faster BSG  wind, which produced a short-lived radio emission when hit by the SN shock \citep{che87}.
The radio emission re-appeared above   noise $\sim1200$ days after the explosion \citep{sta92}, and has been  monitored with the Australia Telescope
Compact Array (ATCA) since then.
\citet{gae97} calculated that, 
between the SN event and the second radio turn-on in mid-1990, the radio emitting regions expanded at an average velocity of $\sim35,000$ km s$^{-1}$, which is possible in a very low-density  inner CSM such as that surrounding Type Ib/Ic SNe.
The expansion velocity sharply decreased to $v_{s}\lesssim7000$ km s$^{-1}$ from day $\sim$1200 \citep{gae97}, when
the radio emission started to increase \citep{bal01}. 
Since day $\sim5000$ the radio emission has been rising at an exponential rate (\citealt{zan10}, Figure  \ref{fig:monitoring}). 
A new phase of the SN$-$CSM interaction is observed at all frequencies after day $\sim8000$ (Figure  \ref{fig:monitoring}; \citealt{zan14b}), when the shock has likely become engulfed in the densest CSM in the ER,
while still expanding above and below the ring.

%
%

%
%
\begin{figure*}[htp]
	 \begin{center}
		\vspace{-1mm}
		\advance\leftskip-0mm
		\includegraphics[trim=0mm 0.0mm 0.0mm 0mm, clip=true,width=119mm, angle=0]{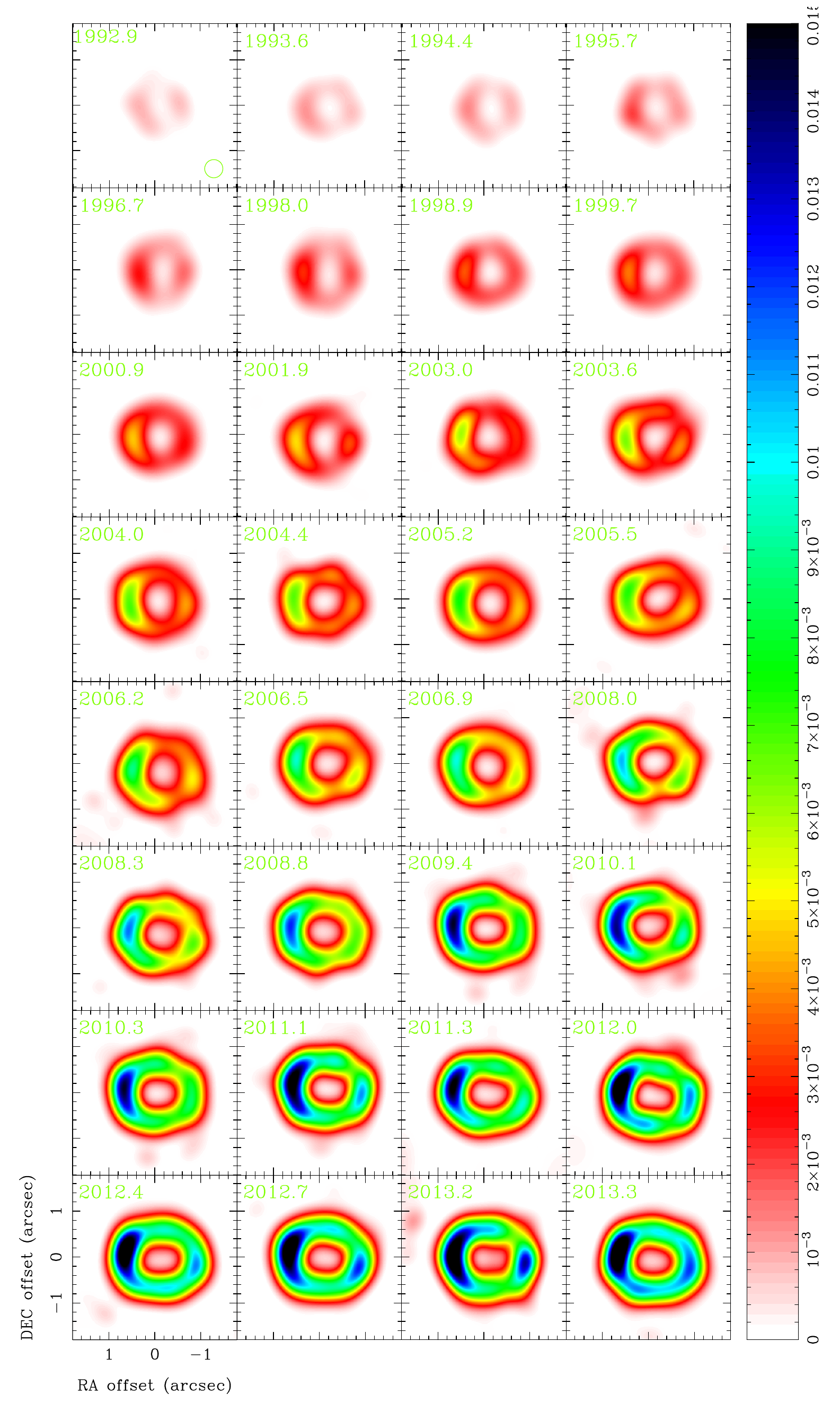}
		\caption{(Caption on the following page.)}
		\label{fig:df}
	\end{center}
\end{figure*}
\begin{figure*}[t]
	\begin{center}
		\contcaption{9 GHz images of SNR 1987A from 1992 to 2013, adapted from \cite{ng13} (see Figure 1 
		therein). Images include data from 
		\citet{gae97}, \citet{ng08}, and \citet{ng13}. Colour-bar units are Jy beam$^{-1}$.
		Credit: \citet{ng13}, \citet{zan14b}.
		\vspace{-0mm}
		}
		\label{Intro:fig_3cm}
	\end{center}
\end{figure*}

\vspace{-2mm}
\subsection{Morphology} 
\label{morph}

\begin{figure*}[!t]
\begin{center}
\vspace{-0mm}
\advance\leftskip-8.5mm
\includegraphics[trim=0mm 0.0mm 0.0mm 0mm, clip=true,width=101.0mm, angle=0]{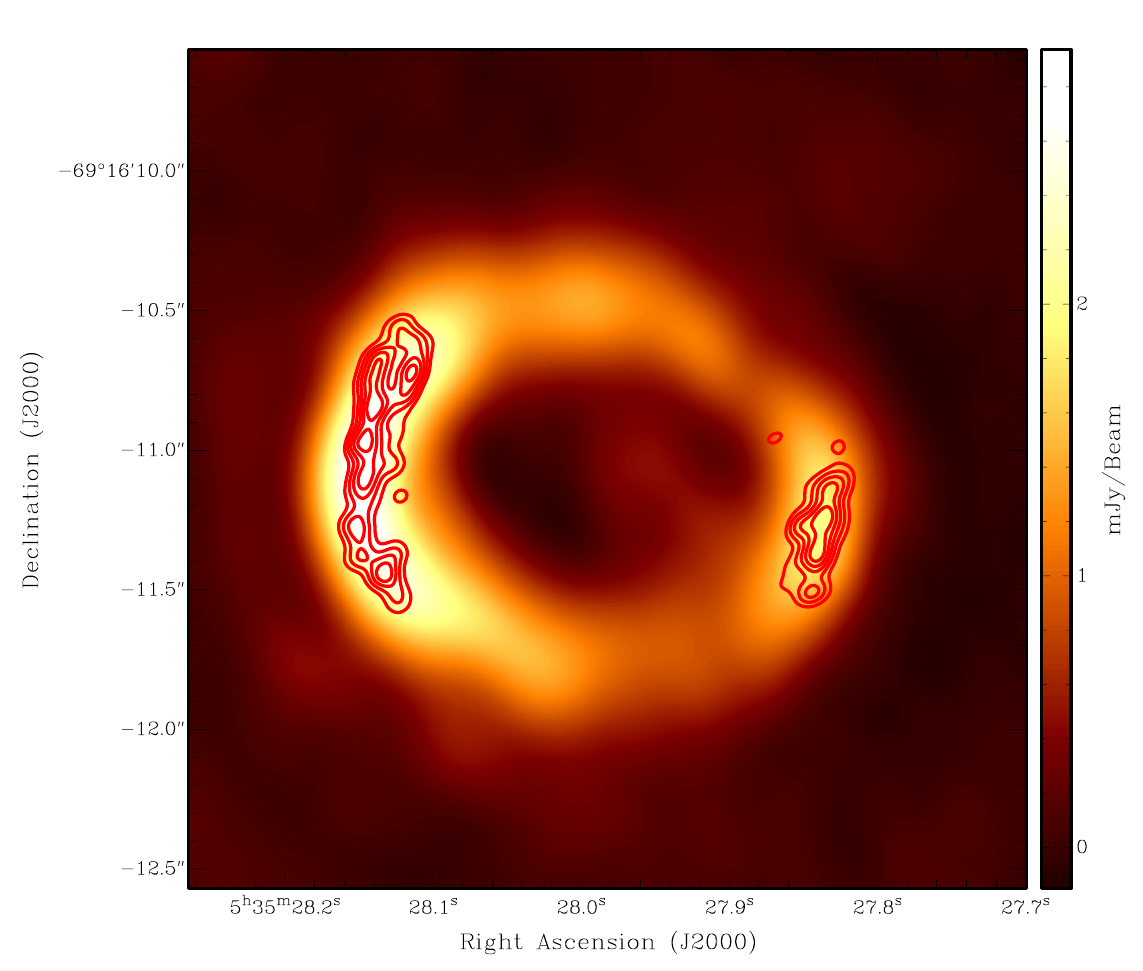}
\end{center}
\vspace{-2.0mm}
\caption{\footnotesize{
Contours of the 2010 VLBI image of SN 1987A at 1.4 GHz are overlaid with the 44 GHz image derived from 2011 observations \citep{zan13}. The 44 GHz image has a resolution of $0\farcs2$, while the VLBI image has a restoring beam of $61\times 95$ mas. The VLBI observations have been instrumental to align images prior to the derivation of spectral index maps (see \citealt{zan14,zan13}).
Credit: \citet{zan14b}.
\\} 
\label{fig:44-LBA}}
\end{figure*}

The morphology of the non-thermal radiation emitted by relativistic electrons accelerated in the remnant has been monitored with the ATCA since 1992 
using images at 9 GHz \citep{sta92}, with a spatial resolution of 0\farcs5 (\citealp{gae97,ng08}).  These images have provided insight into the 
trademarks of the radio emission (see Figure~\ref{Intro:fig_3cm}; \citealt{ng13}), such as: 
(\textit{a}) the steady brightening; 
(\textit{b}) the ring-like structure evolved from a two-lobe morphology; 
(\textit{c}) the east-west asymmetry, with brightness peak on the eastern lobe. 
As discussed by \citet{ng13} (left plots in Figure  \ref{fig:monitoring}), by day $\sim$8000 the faster eastbound shocks (\citealt{zan13}) have possibly overtaken 
the ER. This would have resulted in a reduction of the radio emission from the eastern
rim and, thus, impacted the overall brightness asymmetry (see Figure \ref{fig:monitoring}).
High-resolution images obtained at 44 GHz with the ATCA, and at 1.4 GHz via very-long baseline interferometry (VLBI)  with the Australian Large Baseline Array  (LBA; see Figure \ref{fig:44-LBA}; \citealt{zan14b}), have shown small-scale structures more prominent on the eastern lobe, which are likely due to the shock wave interacting with discrete clumps near the inner surface of the SN shell (see \citealt{ng11}).

%
%

%
%
\begin{figure*}[!t]
\begin{center}
\vspace{-1mm}
\advance\leftskip-8.5mm
\includegraphics[trim=0mm 0.0mm 0.0mm 0mm, clip=true,width=151.0mm, angle=0]{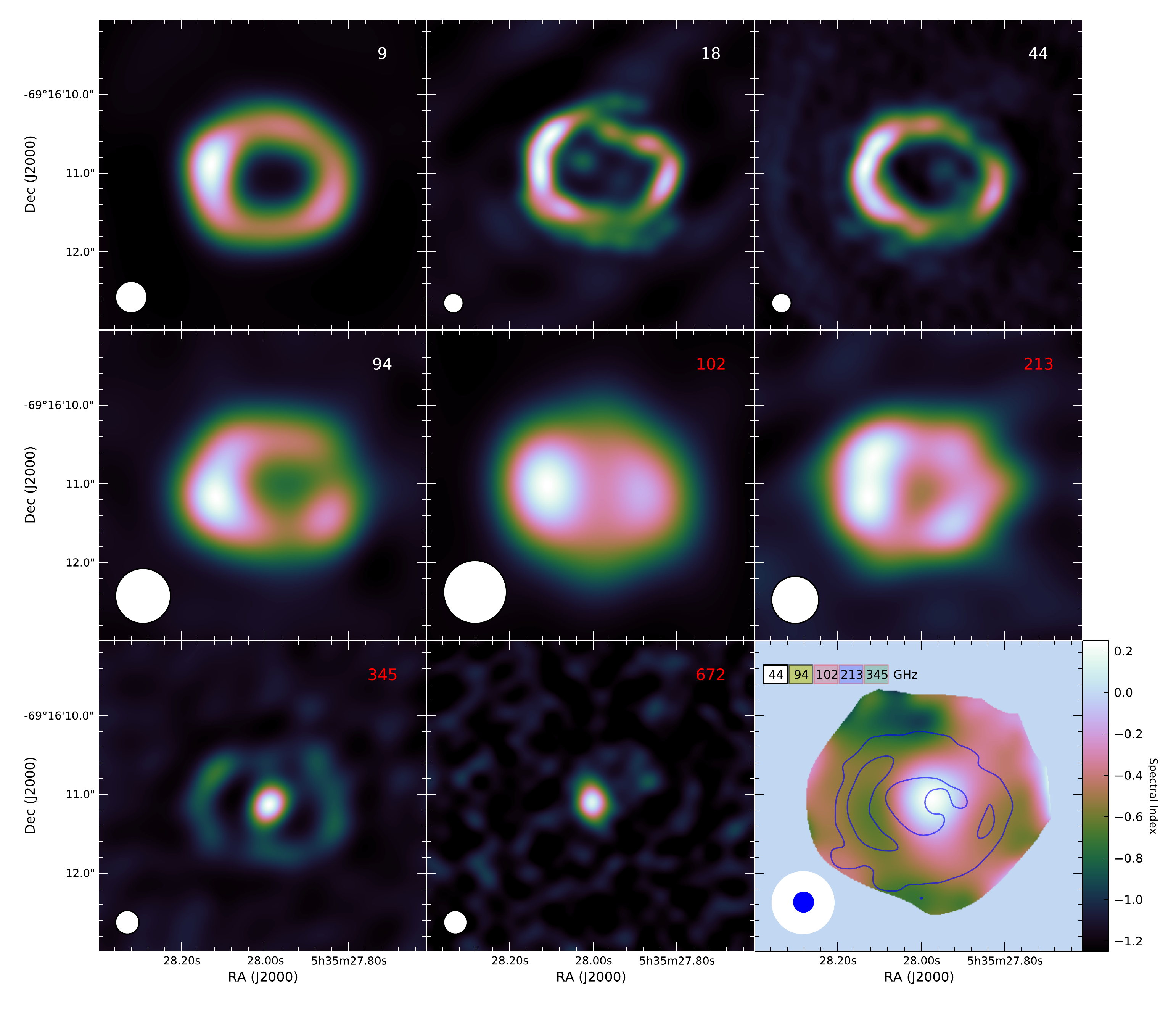}
\end{center}
\vspace{-3.0mm}
\caption{\footnotesize{
Mosaic of radio images of the remnant of SN 1987A at day 9280.  Each image frequency is indicated on the top right corner, in GHz.  
Images below 100 GHz are derived from ATCA observations (\citealp{ng13,zan13,lak12a}), images above 100 GHz are from ALMA data \citep{zan14}. All images have been super-resolved with a circular beam (lower left corner). 
The bottom right image is the spectral index map obtained by combining images at five frequencies, i.e. 44, 94, 102, 213, and 345 GHz (see \citealt{zan14}). The spectral image is overlaid with contours (in blue) of the $15\%$ and $60\%$ emission levels at 44 GHz \citep{zan13}. Credit: \citet{zan14b}.
\\} 
\label{fig:Maps}}
\end{figure*}

\subsection{Asymmetry} 
\label{asym}

The remnant asymmetry 
has been confirmed in all observations from radio to sub-mm (see Figure \ref{fig:Maps}) and discussed for observations at contemporaneous epochs. The east-west asymmetry ratio is 1.4$-$1.5 at 18 and 44 GHz \citep{zan13}, and it shows a gradual decrease in images from 44 to 345 GHz \citep{zan14}, where there are hints of a reversed trend. As the synchrotron lifetime becomes shorter at higher frequencies, there would be regions in the eastern side of the remnant where electrons might be unable to cross the emission sites within their radiative life-time. In this scenario, the synchrotron emission would require the presence of relatively fresh injected and/or re-accelerated electrons to match the emission distribution at lower frequencies. \citet{zan14} estimated that, at frequencies higher than 213 GHz, the electrons might be unable to cross the emission regions on the eastern lobe.

Further, the asymmetry in the shock  expansion velocities, with significantly faster eastbound shocks \citep{zan13},  has been related to  an asymmetric explosion (\citealp{zan13,zan14}).
The possibility of an asymmetric initial explosion was suggested by \citet{che89}, while \citet{pod07} 
proposed a binary merger to reproduce the BSG progenitor and the rings formed from the BSG$-$RSG wind interaction.
Although marked asymmetries have also been seen in the inner ejecta (\citealp{kja10, lar13}), the recent detection of $^{44}$Ti emission with the Nuclear Spectroscopic Telescope Array (\textit{NuSTAR}) provides direct evidence
of large-scale asymmetry in the explosion \citep{bog15}.

%
%
\begin{figure*}[!t]
\begin{center}
\vspace{0mm}
\advance\leftskip-11mm
 \includegraphics[trim=0mm 0.0mm 0.0mm 0mm, clip=true,width=146mm, angle=0]{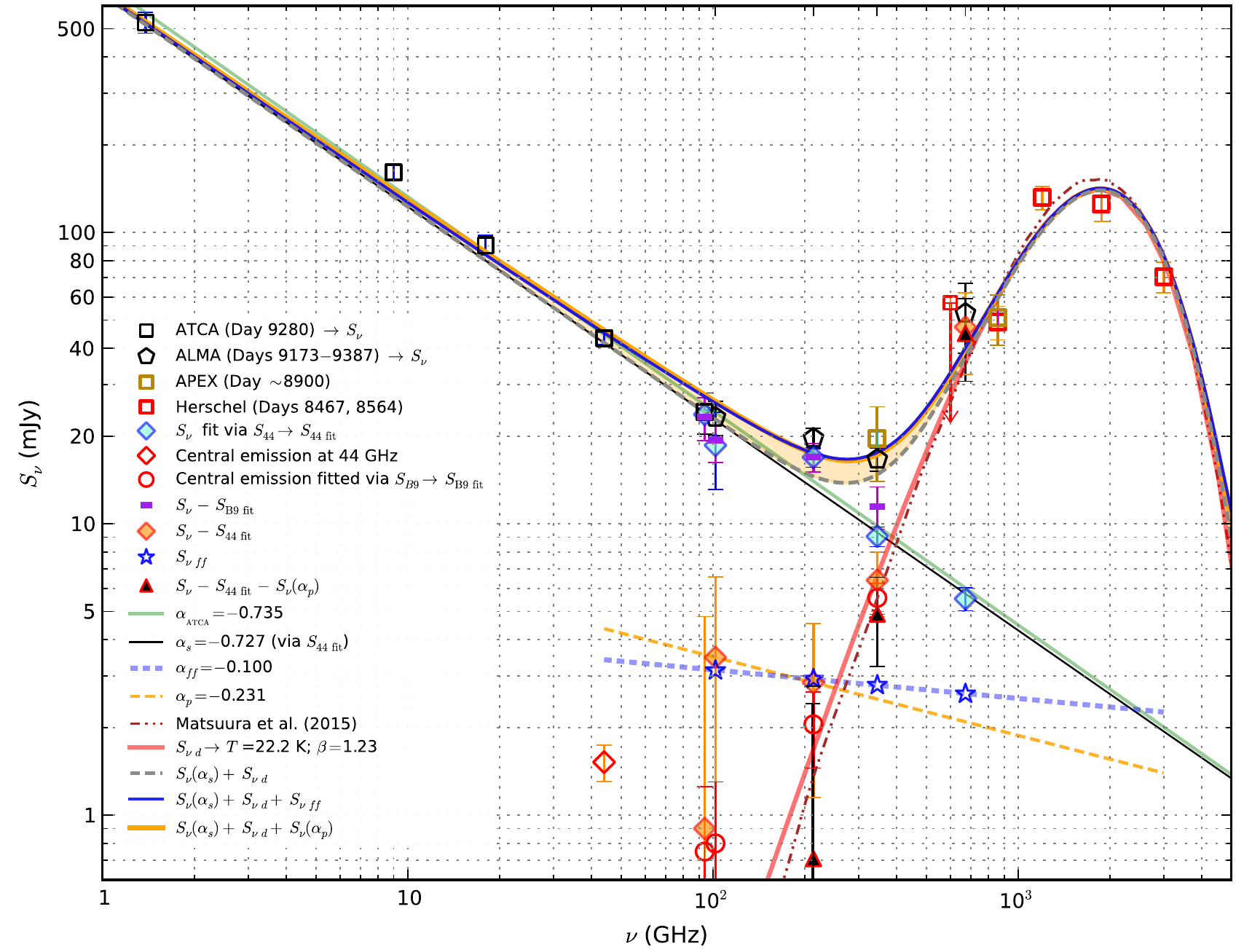}
\vspace{-5mm}
\caption{\footnotesize{
SED
of SNR 1987A from radio to FIR, with data from: ATCA at 1.4 GHz \citep{zan14b}, 
9 GHz \citep{ng13}, 18 and 44 GHz \citep{zan13},  and 94 GHz \citep{lak12a}; ALMA  \citep{zan14}; 
 the Atacama Pathfinder EXperiment (APEX) at 345 and  857 GHz \citep{lak12b}; the Herschel Space Observatory at 600$-$3000 GHz \citep{mat11}. 
The brown dash-dot-dotted curve is the  amorphous carbon dust fit for ALMA data and \textit{Herschel} observations carried out in 2012 \citep{mat15}.
To match the average epoch of ALMA observations, ATCA data are scaled to day 9280, via  exponential fitting parameters derived for ATCA flux densities from day 8000, as measured at 1.4, 8.6 and 9 GHz \citep{zan14b}. 
The hollow red diamond indicates the central emission measured at 44 GHz as reported by \citet{zan13} scaled to day 9280.
 The difference between possible spectral fits is highlighted in light orange. Credit: \citet{zan14}.}
 \label{fig:SED}}
\end{center}
\end{figure*} 

\vspace{-1mm}
\subsection{Spectral index} 
\label{SED}

A progressive hardening of the radio spectra obtained from ATCA multi-frequency flux monitoring has been observed since day $\sim$2500 \citep{zan10}, and up to day $\sim$8000. The results have pointed to sub-DSA particle transport \citep{kir96}, which leads to increasingly efficient particle acceleration and cosmic ray production.
 On a local scale, high-resolution spectral maps have consistently shown 
steeper synchrotron spectral indices and, thus, lower compression ratios, mainly localised on the  eastern lobe of the remnant (see Figure~\ref{fig:Maps}; \citealt{zan13,zan14}).
A softer particle spectrum on the eastern regions  
and the higher Mach numbers measured for the eastbound shocks \citep{zan13}, are likely to create the conditions for  magnetic field amplification \citep{bel01} with efficient cosmic ray production. 
Local magnetic field amplification is compatible with significant synchrotron losses at higher frequencies, which can explain 
the decrease of the east-west asymmetry ratio  of the emission in the transition from radio to sub-millimetre wavelenghts  \citep{zan14}.

The spectral energy distribution (SED) from 1.4 to 3000 GHz (see Figure~\ref{fig:SED}) for 2011$-$2012 data yields a `global' synchrotron spectral index $\alpha=-0.73$, which is confirmed in observations at low frequencies (72$-$230 MHz) carried out with the Murchison Widefield Array (MWA) in late 2013 \citep{cal16}. The absence of a spectral turnover down to 72 MHz, places an upper limit of $n_e=110$ cm$^{-3}$ to the electron density  of the medium in which the SN radio emission propagates \citep{cal16}.

%
%
\begin{figure}[!tp] 
\begin{center}
\vspace{1mm}
\advance\leftskip-0.0mm
\includegraphics[trim=0mm -8mm 1.0mm 0.0mm, clip=true,width=100.55mm, angle=0]{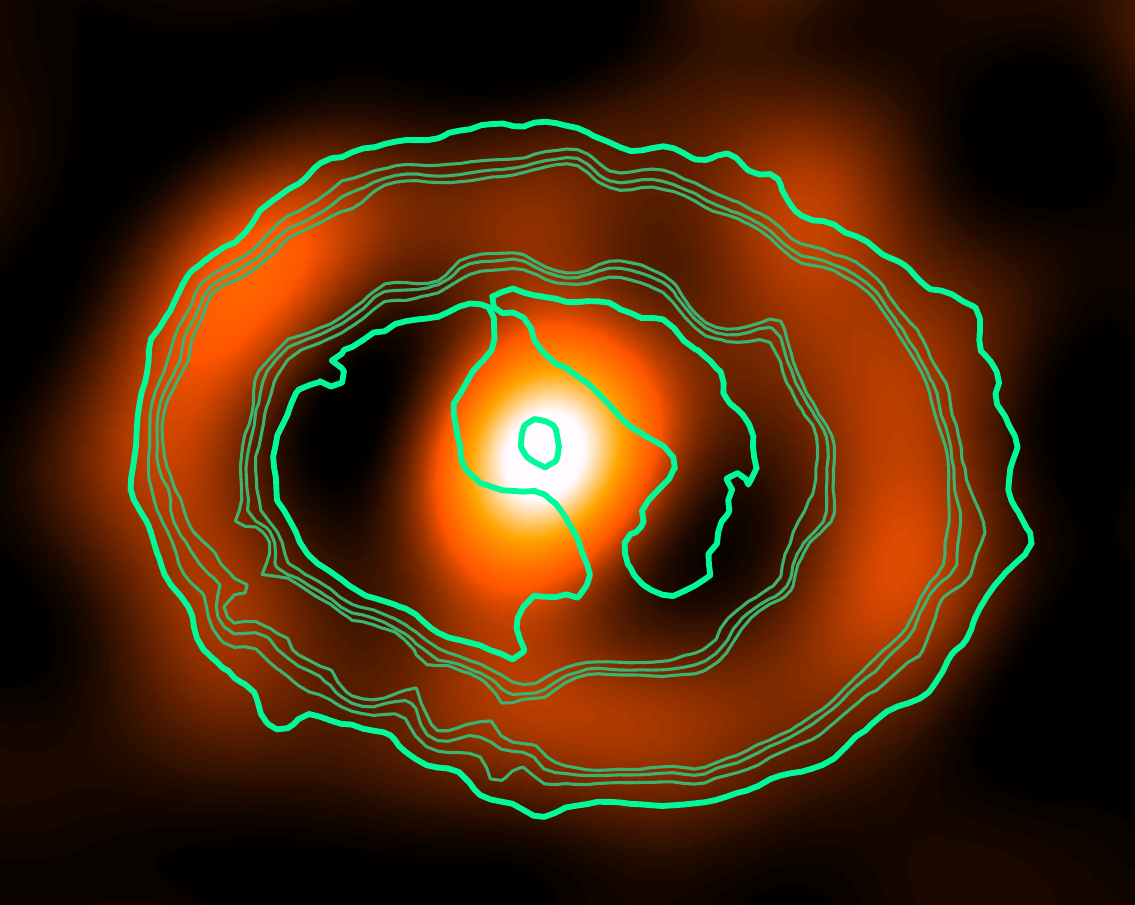}
\vspace{-7.0mm}
\caption[Caption]{
An outline of the equatorial ring and inner debris, as seen with \textit{HST} (green/blue contours), on top of ALMA observations of the remnant at 345 GHz \citep{zan14} (red/orange). 
The dust in the central region is rendered. 
(Hubble data: NASA/STScI/CfA/P. Challis and R. Kirshner.) Credit: \citet{zan14b}.}
\label{fig:ALMA_01}
\end{center}
\end{figure} 

\vspace{-5mm}
\section{Compact remnant}
\label{ns}

The search for a compact object in the inner regions of SNR 1987A has been undertaken on multiple fronts.
It has been shown that flat spectral synchrotron indices can be associated either with accretion onto a black hole or with a pulsar wind nebula (PWN; \citealt{gae06}). Spectral index maps of the remnant derived at different frequencies and epochs have consistently featured a flat central component \citep{pot09,lak12a,zan13,zan14}. 

The discovery with ALMA \citep{ind14,kam13} and \textit{Herschel} \citep{mat11,mat15} of  $\sim0.8\,\msun$ of newly-forming dust grains  in the SNR interior, complicates further the detection of a PWN expected to be expanding into the inner ejecta. The dust sits inside the ER and overlaps with the ejecta as seen with the \textit{Hubble Space Telescope} (\textit{HST}; Figure \ref{fig:ALMA_01}). 
Combined ALMA and ATCA observations have allowed one to separate the thermal and non-thermal components of the SNR emission \citep{ind14,zan14}. After subtraction of the dust emission and the synchrotron emission from the outer shock, the residual emission of  $\sim3$ mJy (see Figure \ref{fig:SED}) in the central region  could correspond to the synchrotron emission from an embedded pulsar, possibly located at $\sim50$ mas from the SN site.
This PWN position would match the centre of the inner emission detected at 44 GHz (Figure~\ref{fig:44-LBA}), and would imply a natal kick of $\sim400$ km s$^{-1}$, for a possible pulsar period of $\sim$150 ms  \citep{zan14}. An asymmetric explosion of the progenitor could have led to a high-velocity neutron star  \citep{jan12} and, thus, to the formation of a PWN offset from the SN site. 

After unsuccessful searches for the pulsar signal with the Parkes telescope \citep{man88,man07}, the preliminary analysis of data from an extended pulsar search campaign in 2013 has tentatively identified several candidates with period $30\lesssim\ P \lesssim 300$ ms \citep{zan14b}. However, there is yet no convincing detection and the search continues.

\end{document}